\documentclass[letter,conference,10pt]{IEEEtran}

\mathchardef\Gamma="0100 \mathchardef\Delta="0101
\mathchardef\Theta="0102 \mathchardef\Lambda="0103
\mathchardef\Xi="0104 \mathchardef\Pi="0105
\mathchardef\Sigma="0106 \mathchardef\Upsilon="0107
\mathchardef\Phi="0108 \mathchardef\Psi="0109
\mathchardef\Omega="010A

\newcommand{\outline}[1]{}

\usepackage{xspace}
\usepackage{url}
\usepackage{graphicx}
\usepackage{latexsym}
\usepackage{amssymb}
\usepackage{amsfonts}
\usepackage{psfrag}
\usepackage{wrapfig}
\usepackage{comment}
\usepackage{color, colortbl}

\newcommand{\Comment}[1]{}

\usepackage{booktabs} 
\usepackage{listings}
\usepackage[bookmarks=false,unicode]{hyperref}
\newcommand{\theTitle}{Are Today's SDN Controllers Ready for Primetime?}
\newcommand{\theAuthor}{Stephen Mallon, Vincent Gramoli and Guillaume Jourjon}
\newcommand{\theKeywords}{Many-core Architecture, Software Defined Networks}
\usepackage{subfigure}
\hypersetup{
  colorlinks=true, linkcolor=black, citecolor=black, urlcolor=blue,  
  pdftitle={\theTitle}, pdfauthor={\theAuthor}, pdfkeywords={\theKeywords}
}
\usepackage{epstopdf}
\usepackage{multicol}
\usepackage[nolist,nohyperlinks]{acronym}
\usepackage{amssymb}
\usepackage{url}
\usepackage{wrapfig}
\usepackage{xcolor}
\usepackage{xspace}
\usepackage{tikz}
\usepackage{flushend}

\usepackage[figuresright]{rotating}
\newenvironment{smallenum}{
\begin{enumerate}
    \setlength{\topsep}{-1pt} 
    \setlength{\partopsep}{-1pt}
  \setlength{\itemsep}{0pt}
  \setlength{\parskip}{-1pt}
  \setlength{\parsep}{-1pt}
}{\end{enumerate}}

\newcommand{\vincent}[1]{} 
\newcommand{\opfont}[1]{\texttt{#1}}

\title{Are Today's SDN Controllers Ready for Primetime?}%

\author{\IEEEauthorblockN{Stephen Mallon\IEEEauthorrefmark{2},
Vincent Gramoli\IEEEauthorrefmark{1}\IEEEauthorrefmark{2}, and
Guillaume Jourjon\IEEEauthorrefmark{1}}
\IEEEauthorblockA{\IEEEauthorrefmark{1} Data61-CSIRO, Australian Technology Park, Eveleigh, NSW, Australia\\
guillaume.jourjon@data61.csiro.au}
\IEEEauthorblockA{\IEEEauthorrefmark{2} University of Sydney, Sydney, NSW, Australia.\\
smal8373@uni.sydney.edu.au, vincent.gramoli@sydney.edu.au}}

\begin{document}
\maketitle

\begin{abstract}
SDN efficiency is driven by the ability of controllers to process small packets based on a global view of the network. The goal of such controllers is thus to treat new flows coming from hundreds of switches in a timely fashion. In this paper, we show this ideal remains impossible
through the most extensive evaluation of SDN controllers.
We evaluated five state-of-the-art 
SDN controllers and discovered that the most efficient one spends a fifth of his time in packet serialization.
More dramatically, we show that this limitation is inherent to the object oriented design principle of these controllers. They all treat each single packet as an individual object, a limitation that induces an unaffordable per-packet overhead. To eliminate the responsibility of the hardware from our results, we ported these controllers on a network-efficient architecture, Tilera, and showed even worse performance. We thus argue for an in-depth rethinking of the design of the SDN controller into a lower level software that leverages both operating system optimizations and modern hardware features. 

\end{abstract}


\section{Introduction}
\label{sec:introduction}

With the advent of SDN, the control of networks has migrated from routers with dedicated hardware support to more flexible software solutions running on general purpose platforms.
This shift raises a new issue due to the inadequacy between these controller applications 
and their underlying concurrent architectures.

An SDN~\cite{MABP+08} consists of many switches requesting forwarding decisions 
from a centralized controller based on traffic flow headers.  This centralization at the controller allows 
to simplify network decisions based on the global view of the network map, but it also induces a 
high traffic rate due to many switches sending new flow packets to the same 
controller.
For example, 
given that 20\% of new traffic flows per switches have a 10\,$\mu{}$s inter-arrival rate in existing datacenters~\cite{benson:2010}, a single controller
should treat 20 millions flow requests per second to control 200 switches, a traffic that no existing controllers can support. 
Furthermore, the size of these packets, typically very small, 
exacerbates the bottleneck problem of the controller networking stack.
In fact, as the ingress traffic rate increases the per-packet overhead, 
induced by interrupts and memory accesses,
predominates rapidly the latency at which the controller processes packets. 

Datacenters predominantly exploit general purpose multi-core processors  for various 
services and most SDN controllers were naturally implemented on these common platforms.
The complexity of the cores sitting on these processors induce however a high power consumption
with limited improvements in performance, as imposed by the Pollack's Rule~\cite{borkar:2011}.

\sloppy{Many-core processors attempt to address this problem by using a large number of simpler cores
that communicate using a Network-on-Chip (NoC) with limited or no support for cache coherence~\cite{borkar:2011}. 
Their cores are slower than multi-core processors', but overall many-core processors offer a higher performance over energy ratio than multi-core processors~\cite{GG16}.
Many-core processors like the ones manufactured by Tilera
already proved instrumental in high traffic applications, for example
running the Suricata Intrusion Detection System (IDS) at high throughput~\cite{TileraSuricata:2012} and the Facebook in-memory cache application with unprecedented performance over energy ratio~\cite{berezecki-etal:2011}.
So one could wonder whether SDN controllers were ready to leverage many-cores platforms.
}

In this paper, we perform the most extensive evaluation of SDN controllers to date. To this end, 
we evaluate in detail state-of-the-art SDN controllers: the original OpenFlow controller, NOX~\cite{gude:2008,tootoonchian:2012};  the first controller to leverage concurrency, Maestro~\cite{ngmaestro};
the FloodLight controller that is 
at the heart of a controller in production~\cite{erickson:2012floodlight}; and the controller that experiences, as far as we know, the highest throughput, Beacon~\cite{erickson:2013beacon}. 
In addition, we also evaluated the OpenDaylight SDN controller~\cite{opendaylight} but do not report its performance that are significantly lower than the others.
Using a 10\,Gbps network and multiple instances of the controller benchmark CBench~\cite{sherwoodcbench}, we compare their  
throughput
in millions flows per second,  
their flow latency, their associated performance per Watt and 
their refined profiling.
The fact that today's reactive controllers are not deployable in carrier grade network~\cite{juniper} confirms
the underlying research challenge in this area~\cite{Procera, Reactive}.
To generalize our observations, we ported these controllers on a Tilera many-core server 
similar to the one used by Suricata~\cite{TileraSuricata:2012} and a more common x86 server with 2 Intel Xeon processors.

Overall, we unveil a list of problems that illustrate the inadequacy of SDN controllers to exploit the underlying concurrency of modern multi-/many-core platforms.
Our results were surprising as they are less encouraging than previous controller evaluations that 
showed reasonable performance by exploiting slower network or 
loopback configurations. 
First, on both platforms, all controllers experience IO or data structure bottlenecks that tend to flatten out 
their throughput as the level of concurrency increases.
Second, we uncovered that Beacon achieved impressive performance by responding to CBench switches 
using packet-out rather than the expected flow-mod messages, hence presenting a throughput 
 20\% higher while violating the OpenFlow specification.

The performance results obtained on the many-core platform are even more dramatic as they represent a lower portion of our network capacity than on the multi-core platform.
Moreover, while memcached~\cite{berezecki-etal:2011} achieved unprecedented performance over energy ratio
on Tilera, our power consumption measurements indicate that controllers also fail in exploiting the energy saving of many-core.
In particular, due to their object-oriented design, these controllers inherently treat each packet as a separate object, hence requiring costly memory allocations and copies. These memory accesses translate into a substantial per-packet overhead that becomes unbearable on Tilera where the memory bandwidth is not as high as on Intel.

Previous studies already compared the throughput and response time of NOX, Beacon and Maestro, however, these studies were done on machines from 2 to 8 cores only~\cite{erickson:2013beacon,tootoonchian:2012}. We revisited these results on machines with four times as many hardware threads. In particular, we evaluated performance and energy efficiency without loopback interface by exploiting the network. In addition, we tested different architectures, including manycores and traditional multicores embedding multiple sockets. Finally, we extended our evaluation to controllers used in production~\cite{erickson:2012floodlight,opendaylight}.
In the light of this extensive evaluation, our results differ radically from previous ones: the 
performance of existing controllers do not truly leverage the concurrency level of modern machines. In particular, we identified a bug in a mainstream controller, 
a bottleneck problem in a key data structure and a large per-packet overhead due to object management.

In Section~\ref{sec:legacydesign} we present our evaluation methodology.
In Section~\ref{sec:legacy}, we quantify the mismatch between controllers and multi-/many-core platforms in terms of throughput limitations and high latency, and we show how they fail in leveraging the power efficiency of many-cores.
We provide an in-depth profiling to explain these results in Section~\ref{sec:discussion}.
Finally, 
we present the related work in Section~\ref{sec:related} and conclude in Section~\ref{sec:conclusion}.

\section{Methodology}
In this section, we present the SDN controllers and how we had to adapt the code to benchmark it and to fix some bugs. We also present the controller benchmark, the network and power configurations, and the multi-/many-core platforms used in the evaluation.

\label{sec:legacydesign}

\subsection{Controllers under study}

We evaluate the performance and energy consumption of four state-of-the-art controllers on both 
multi-core and many-core architectures:
(i)~NOX is the original OpenFlow controller~\cite{gude:2008} whose recent version is multi-threaded~\cite{tootoonchian:2012};
(ii)~Maestro was the first controller whose performance started scaling with the number of cores~\cite{ngmaestro}; 
(iii)~Floodlight is the controller at the core of the Big Switch Network commercial product~\cite{erickson:2012floodlight}; and 
(iv)~Beacon is presumably the fastest controller to date~\cite{erickson:2013beacon}. 
We also tested the OpenDaylight SDN controller~\cite{opendaylight} but the performance were too low to provide any insightful comparison so we decided to omit the results from the evaluation.

\begin{table}[t]
\centering
\caption{Environment specifications\label{table:spec}}
\setlength{\tabcolsep}{16pt}
\begin{tabular}{lcc}
\toprule
 & Many-core & Multi-core \\
 \midrule
Processors &TileGx-36 & Xeon E5-2450 \\
Number of processors & 4 & 2 \\
Cores per processors & 36 & 8 (16-way) \\
Clock frequency & 1.2\,GHz & 2.1\,GHz\\
Memory & 128\,GB& 128\,GB\\ 
Kernel & 3.10 & 3.8.0 \\ 
gcc & 4.4.6  & 4.8.1\\
Java & 1.6 & 1.7\\
\bottomrule
\end{tabular}

\end{table}

\subsection{Implementation issues and solutions}
We explain now few problems we addressed before being able to evaluate the controllers properly.\footnote{Original teams have been notified of these problems.} Indeed, these modifications were necessary to ensure the same test was running for every controller (i.e. controller responds with flow-mods messages). 
\begin{smallenum}
\item Beacon has a race condition that is currently unpatched, this can result in corrupted OpenFlow messages that leave CBench in an infinite loop. 
If the OpenFlow header size field is not at least 8 bytes (the size of the header) 
the execution results in an infinite loop. The race condition is in
\texttt{core/io/internal/OFStream.java}.
Write methods are synchronized on the outBuf object, however, this reference is changed if the buffer needs to be resized (in \texttt{OFMessageAsyncStream.java:
	appendMessageToBuf}).
\item Floodlight and Beacon use a convoluted handshake involving more steps than CBench expects, the workaround was simply to use \texttt{-D 1000} to delay sending ``packet-ins'' for 1 second after CBench finishes the initial handshake. If this was not done, these controllers would respond with packet-out messages which are a $3^{rd}$ the size of flow-mods expected messages.
\item Floodlight learning switch implementation requires modifying the field \texttt{LEARNING\_SWITCH\_REVERSE\_FLOW = false} in \texttt{learningswitch/LearningSwitch.java}.
in order to avoid sending 2 flow-mods per packet-in. 
CBench has a sanity check to make sure it does not receive more responses than probes, however, FloodLight responds slowly enough that there are always more probes than responses.
\item Maestro requires modifying \texttt{MAXIMUM\_DIVIDE} in \texttt{sys/Constants.java}.
The default implementation of Maestro has a max thread count of 8. To test performance with more threads, we had to increase it by changing the \texttt{MAXIMUM\_DIVIDE} value.

\end{smallenum}

\begin{figure}[t]
\begin{center}
\subfigure[Multicore configuration]{\hspace{-1em}\includegraphics[scale=0.515]{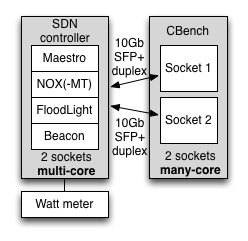}}
\subfigure[Manycore configuration]{\hspace{-1em}\includegraphics[scale=0.515]{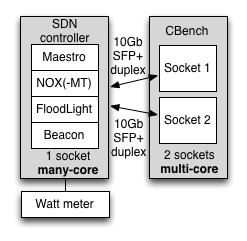}}
\end{center}
\caption{The network and power configurations}
\label{topology}
\end{figure}

\subsection{Multi-core and many-core platforms}\label{ssec:machines}
Our many-core platform is a 1U TILExtreme-Gx36 server featuring 4 TileGx-36 processors, connected to IO modules of $2\times10$\,Gbps SFP+ interfaces and running at 1.2\,GHz.
The cache coherence of each processor can be enabled or disabled but no coherence is 
guaranteed across processors. Each processor embeds 36 cores on the same chip organized into a 
2-dimensional mesh of dies---2 cores per die---communicating through routed messages on a dedicated
Network-on-Chip (NoC). It runs a port of the Linux kernel v3.10, gcc v4.4.6 and Java v1.6, all tuned 
for the Tilera architecture.

Our multi-core platform is a 1U server with 2 Intel Xeon E5-2450 processors connected through a $2\times10$\,Gbps SFP+ Intel X520-DA2 and running at 2.1GHz.
As opposed to the many-core, cache coherence is naturally maintained even across processors.
Each processor embeds 8 hyperthreaded cores for a total of 16 simultaneously supported threads.
It runs Linux Ubuntu 12.04.5 LTS with kernel v3.8.0, gcc v4.8.1 and Java v1.7.
These specifications are summarized in Table~\ref{table:spec}.

\subsection{Testbed setup}
We measure the performance and energy consumption of the four controllers on the two platforms using 
the standard SDN controller benchmark, CBench~\cite{sherwoodcbench} and the Watts up PRO 99333 
Watt meter as depicted in \figurename~\ref{topology}.

CBench emulates OpenFlow switches that it connects to the controller under test.
As opposed to the 4 controllers, CBench is however single threaded,
which makes it the potential performance bottleneck of the experiments as previously observed~\cite{erickson:2013beacon}. 
To overcome this issue we deployed multiple instances of CBench running them on two sockets, each 
socket being connected through independent 10Gbps links to the other machine running the controller.
As the CBench instances do not have to communicate, we run its instances on two sockets in all cases.

CBench
connects
switches to the controller under test.
It emulates switches that generate 82-byte OpenFlow packet-in messages simulating a miss of the switch flow table and a request for a controller's decision.
The controller responds to packet-in messages reactively, creating a flow-mod message containing the OpenFlow header and sent to the switch and cached.
Note that reactive configurations are known to be very difficult to deploy in carrier grade 
networks~\cite{juniper}, and thus remains an interesting research challenge~\cite{Procera, Reactive}.

\begin{figure*}
\hspace{0.5em}
\subfigure[TileGx concurrency\label{fig:final_tilera_threads}]{\includegraphics[width=.32\textwidth]{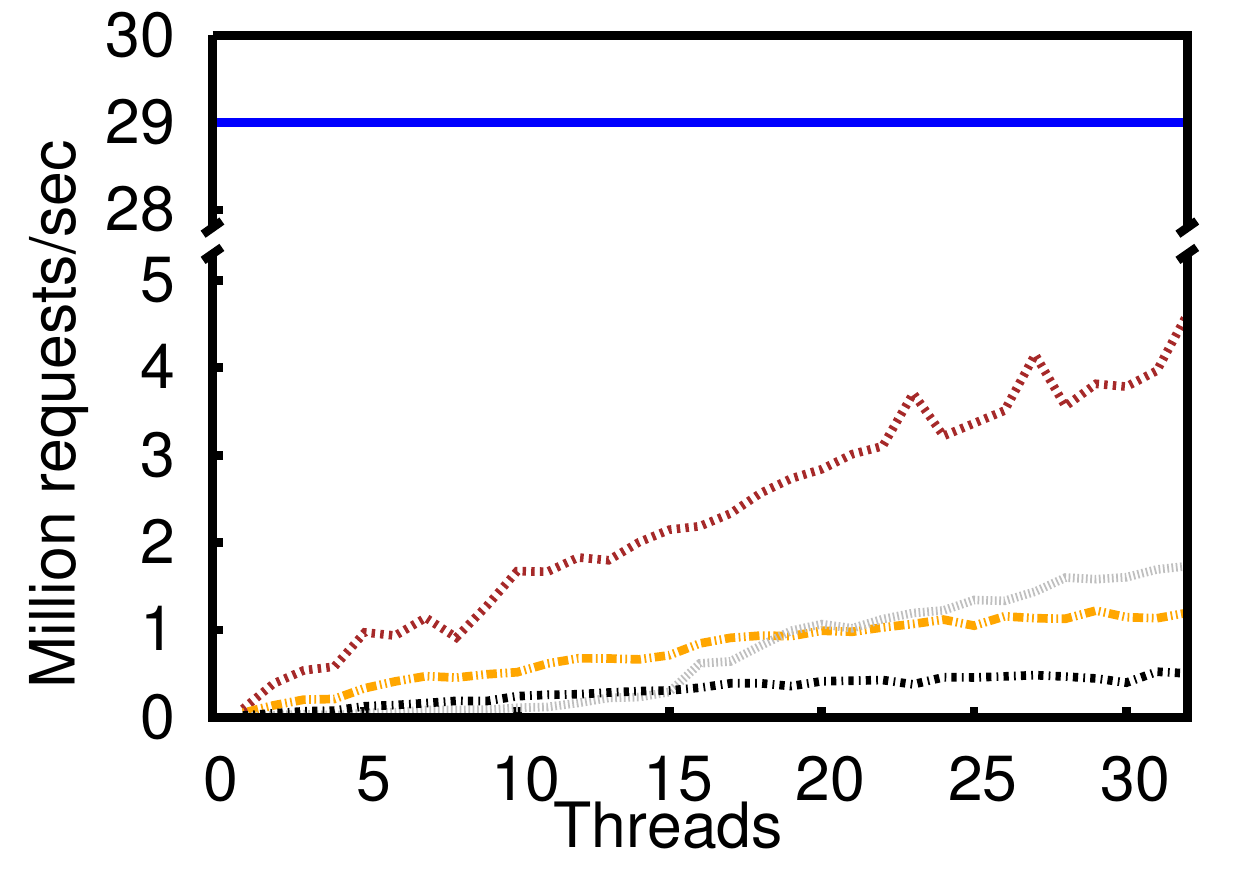}}
\subfigure[TileGx heterogeneity\label{fig:final_tilera_address}]{\includegraphics[width=.32\textwidth]{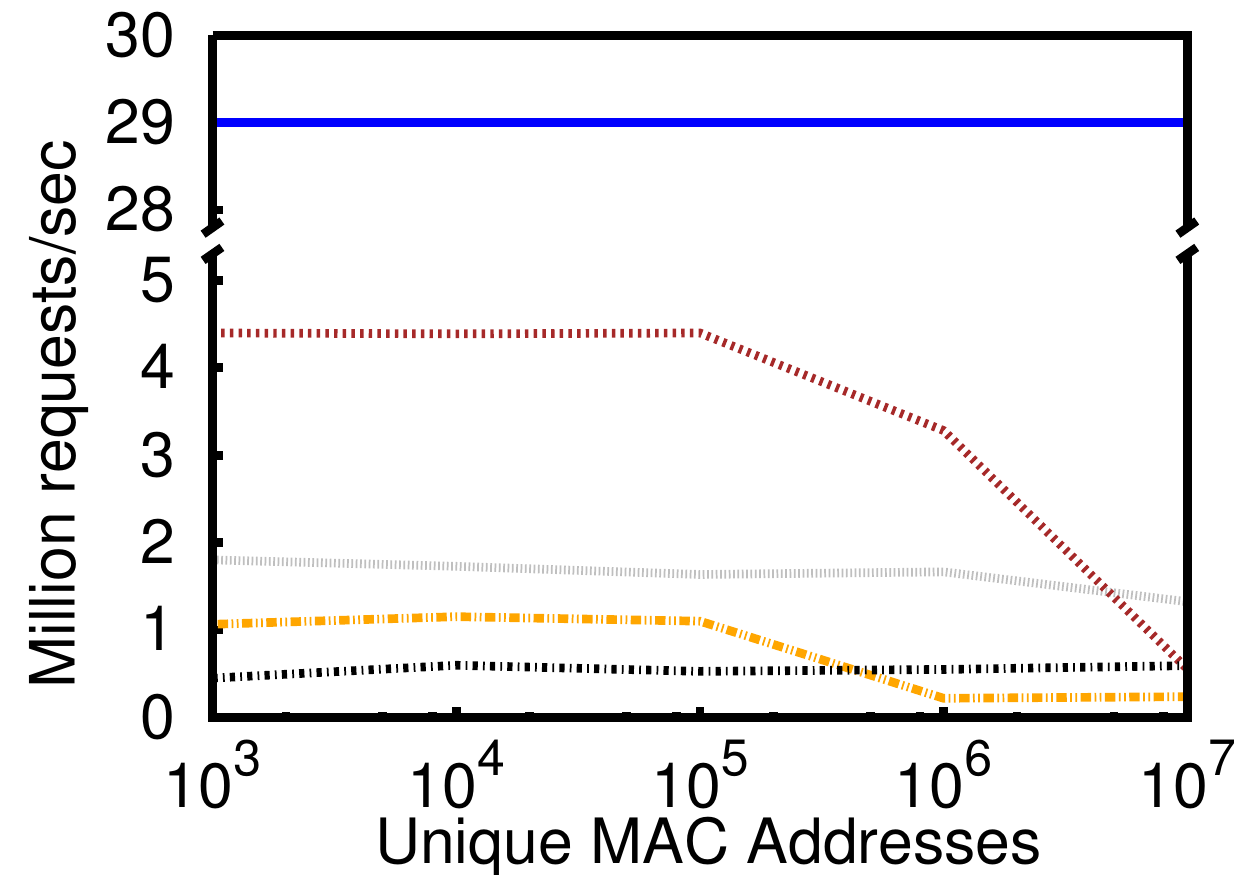}}
\subfigure[TileGx connectivity\label{fig:final_tilera_switches}]{\includegraphics[width=.32\textwidth]{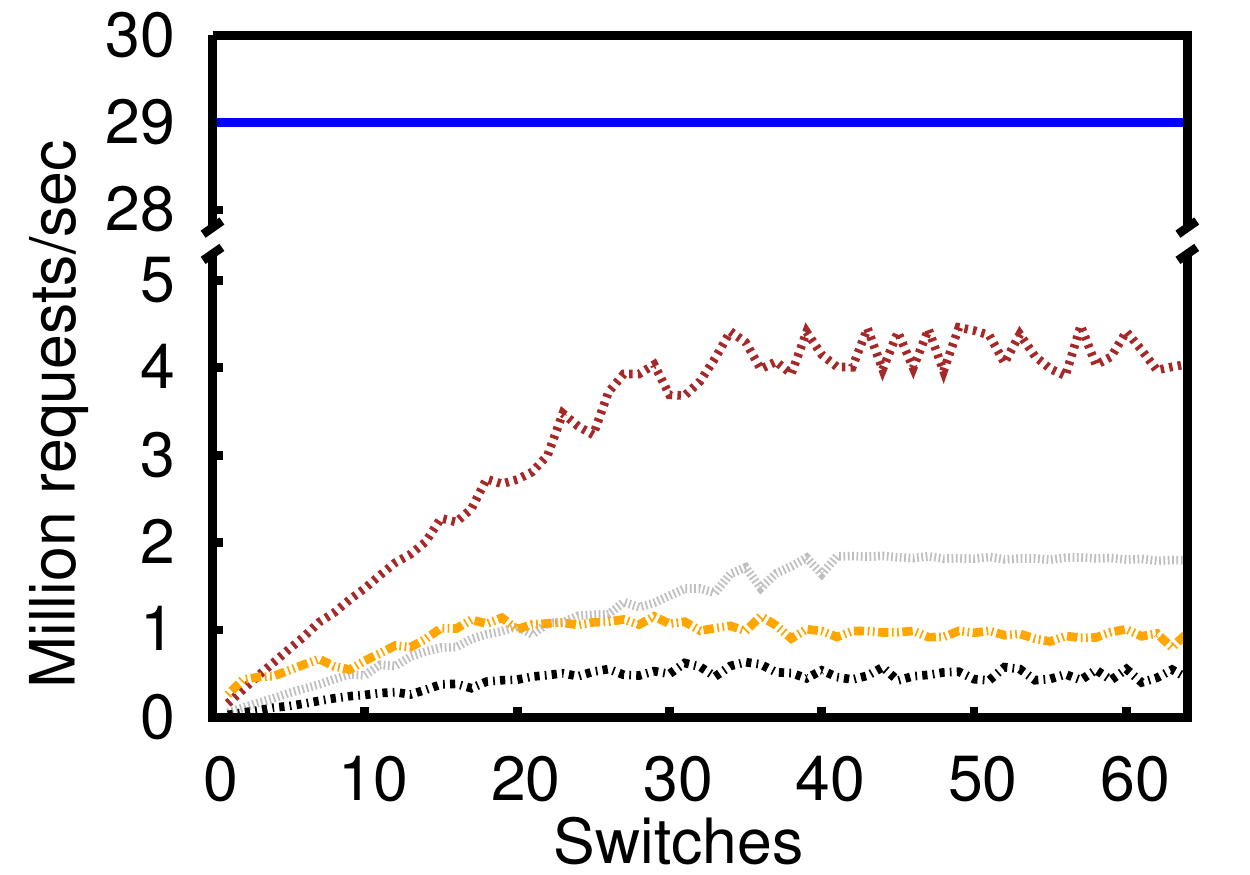}}
\vspace{-1em}
\subfigure[x86 concurency\label{fig:final_x86_threads}]{\includegraphics[width=.33\textwidth]{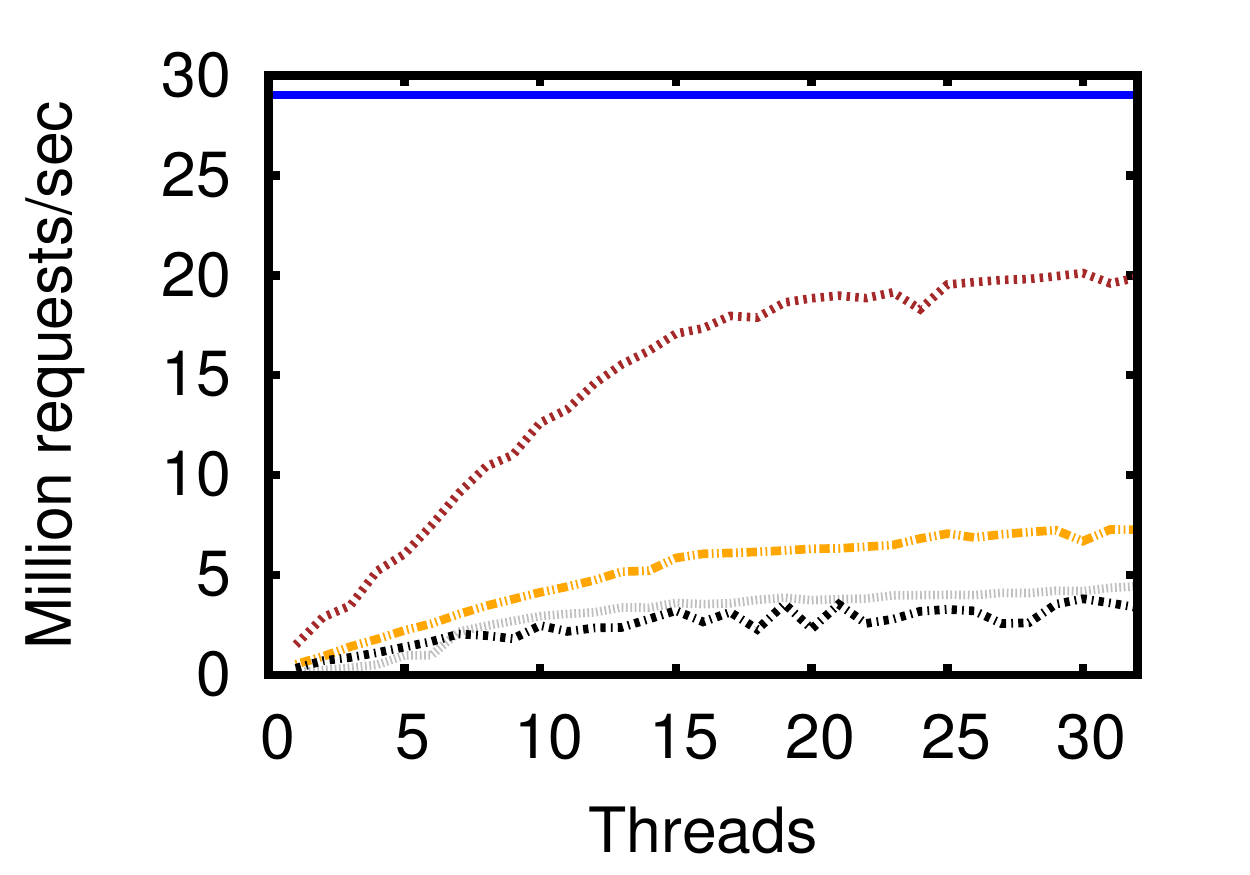}}
\subfigure[x86 heterogeneity\label{fig:final_x86_address}]{\includegraphics[width=.33\textwidth]{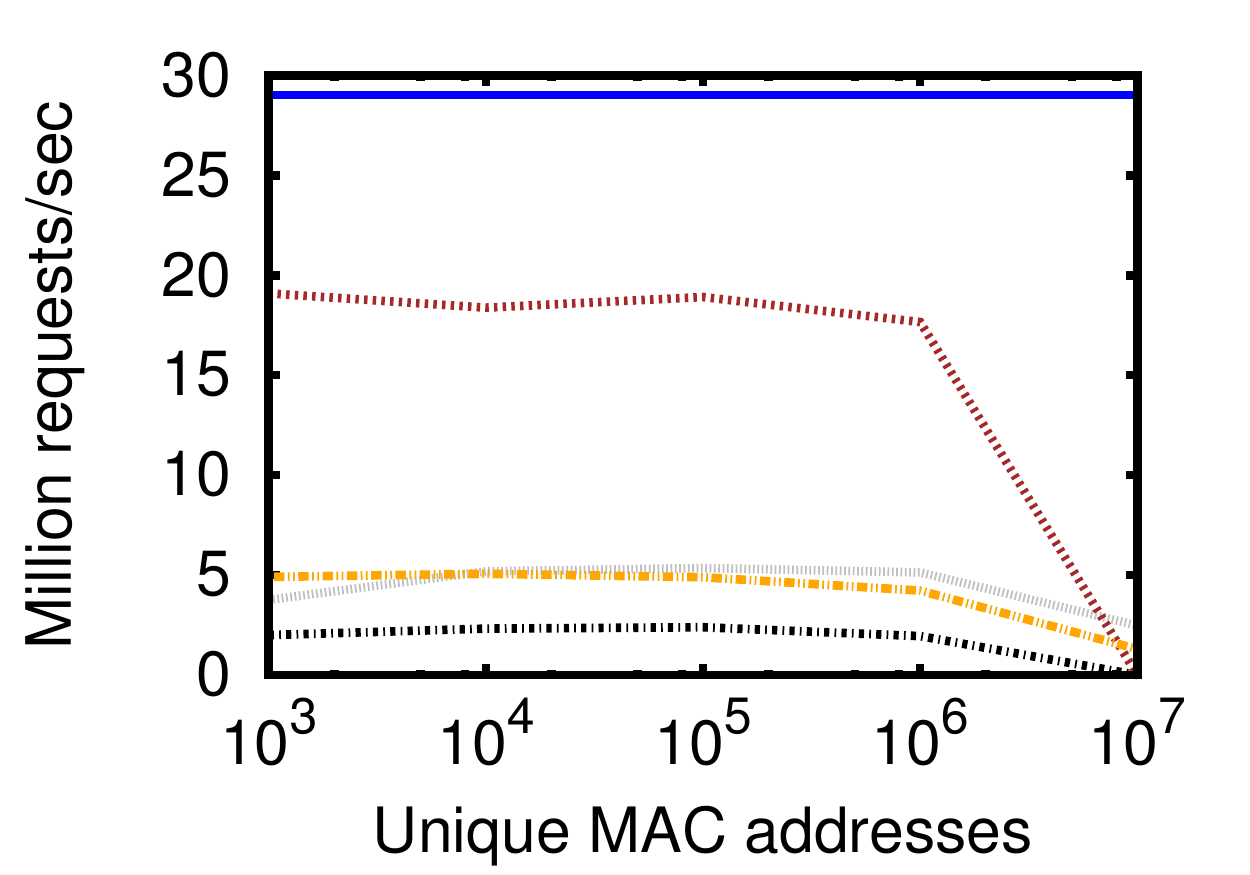}}
\subfigure[x86 connectivity\label{fig:final_x86_switches}]{\includegraphics[width=.33\textwidth]{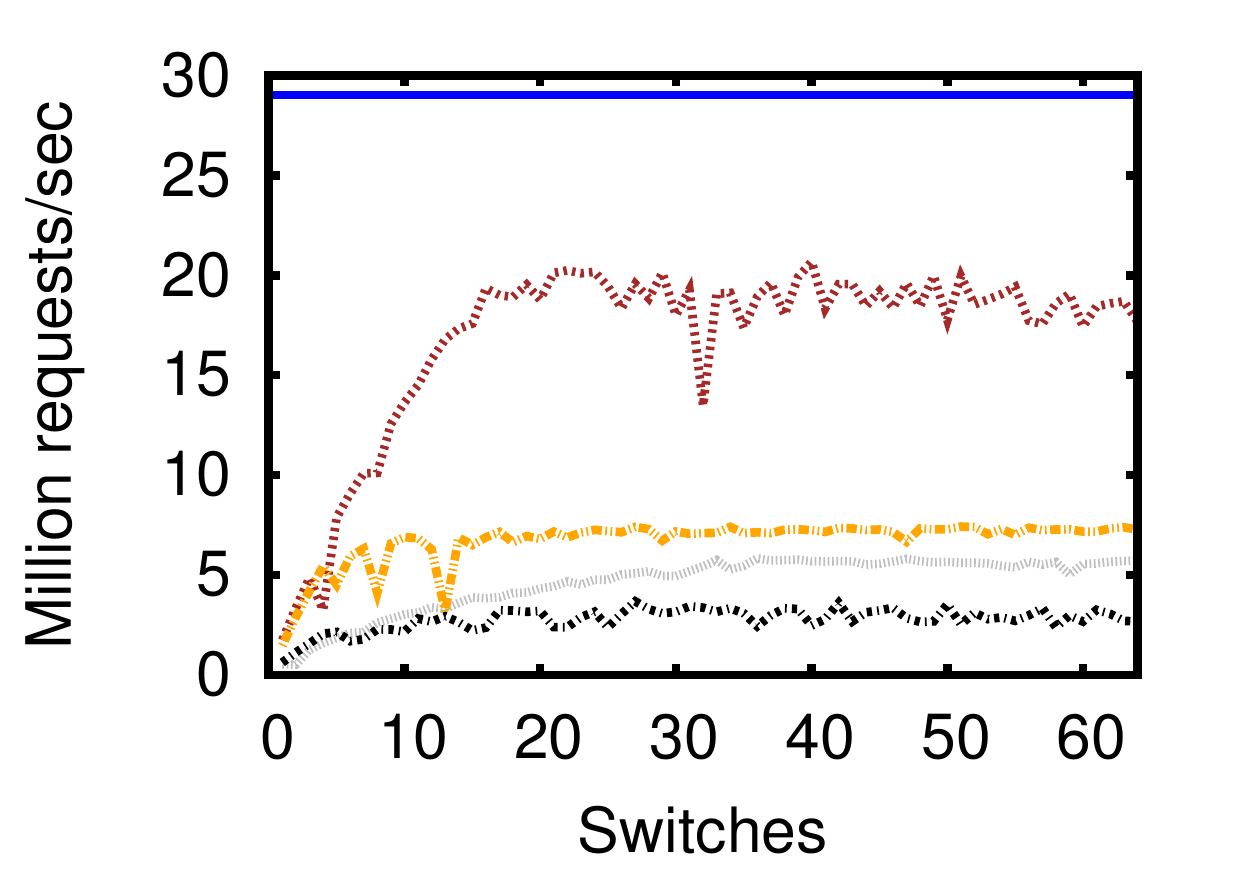}}
\begin{center}
\subfigure{\includegraphics[width=.55\textwidth]{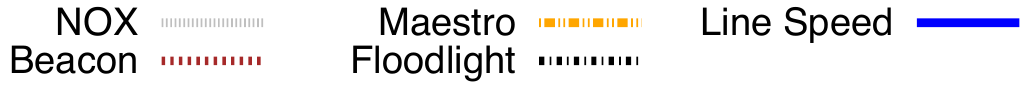}}\\
\end{center}
\caption{Controllers throughput on many-core (top) and multi-core architectures (bottom) as concurrency (left), heterogeneity (center) and connectivity (right) increase}
\label{fig:final_all}
\end{figure*}

\section{Performance Results}
\label{sec:legacy}

This section presents the evaluation of state-of-the-art controllers in terms of maximum number of flows it can handle, latency to respond to these flows and the number of requests per Watts it can achieve. 
We illustrate the concurrency limitations of legacy SDN controllers by comparing the performance on multi-cores and many-cores platforms. 
Each value we plot results from the average obtained from 10 runs of 10 seconds each.

\subsection{Throughput}

We evaluate the throughput of the controllers as the number of flows they can treat per second along 
three dimensions: concurrency, heterogeneity, and scalability.
To test performance under concurrency, we varied the number of threads running the controller from 1 to the number of hardware threads offered by the platform, as detailed in Section~\ref{ssec:machines}.
To test performance under heterogeneity, we varied the number of unique MAC addresses of the flows present in the network. To test scalability, we measured performance as the number of  
switches to connected to the controller increases. 
When not mentioned otherwise, we fixed the number of threads to 32, the number of MAC addresses to $10^6$ and the number of switches to 64.

Figure~\ref{fig:final_all} depicts the throughput on many-core (top) and multi-core (bottom) 
as the concurrency (left), the heterogeneity (center) and connectivity (right) vary.
We generally observe that on the many-core platform, peak performance is substantially lower than line speed (not the change in scale at the top of the $y$-axes of Figures~\ref{fig:final_tilera_threads}-\ref{fig:final_tilera_switches}) and than the performance on the multi-core platform.

On the many-core platform, the controller performance increases with the level of concurrency or the connectivity but the peak throughput (Beacon at 4.8Mreq/sec on Figure~\ref{fig:final_tilera_threads} and 4.4Mreq/sec on Figure~\ref{fig:final_tilera_switches}) is 6$\times$  lower than the line speed of 29Mreq/sec (based on the 1.45Mreq/sec per Gbps of bandwidth~\cite{tootoonchian:2012}). 
On the multi-core platform, no controllers truly exploit all hardware threads. For example, Beacon stops scaling at 18 threads (Figure~\ref{fig:final_x86_threads}) and 17 switches (Figure~\ref{fig:final_x86_switches}). Although this may be due in part to Intel hyperthreading that requires more than 16 threads to start sharing registers, our in-depth profiling shows several bottlenecks detailed in Section~\ref{sec:discussion}. Finally, the results vary as we can observe in Figure~\ref{fig:final_x86_threads} that Beacon almost reaches $20$ Mreq/secs on multi-core while the other controllers only reach a maximum of 7 millions req/secs. 
Finally, Figures~\ref{fig:final_tilera_address} and~\ref{fig:final_x86_address} show the throughput as the number of unique MAC addresses present in the network increases. 
Interestingly, performance of Java-based controllers on both architectures drops at 10 millions unique MAC addresses.

\subsection{Latency}

\begin{figure*}[t]
\centering
\subfigure[TileGx w/ a single switch\label{fig:latency1t}]{\includegraphics[width=.79\columnwidth, clip=true, viewport=28 5 355 250]{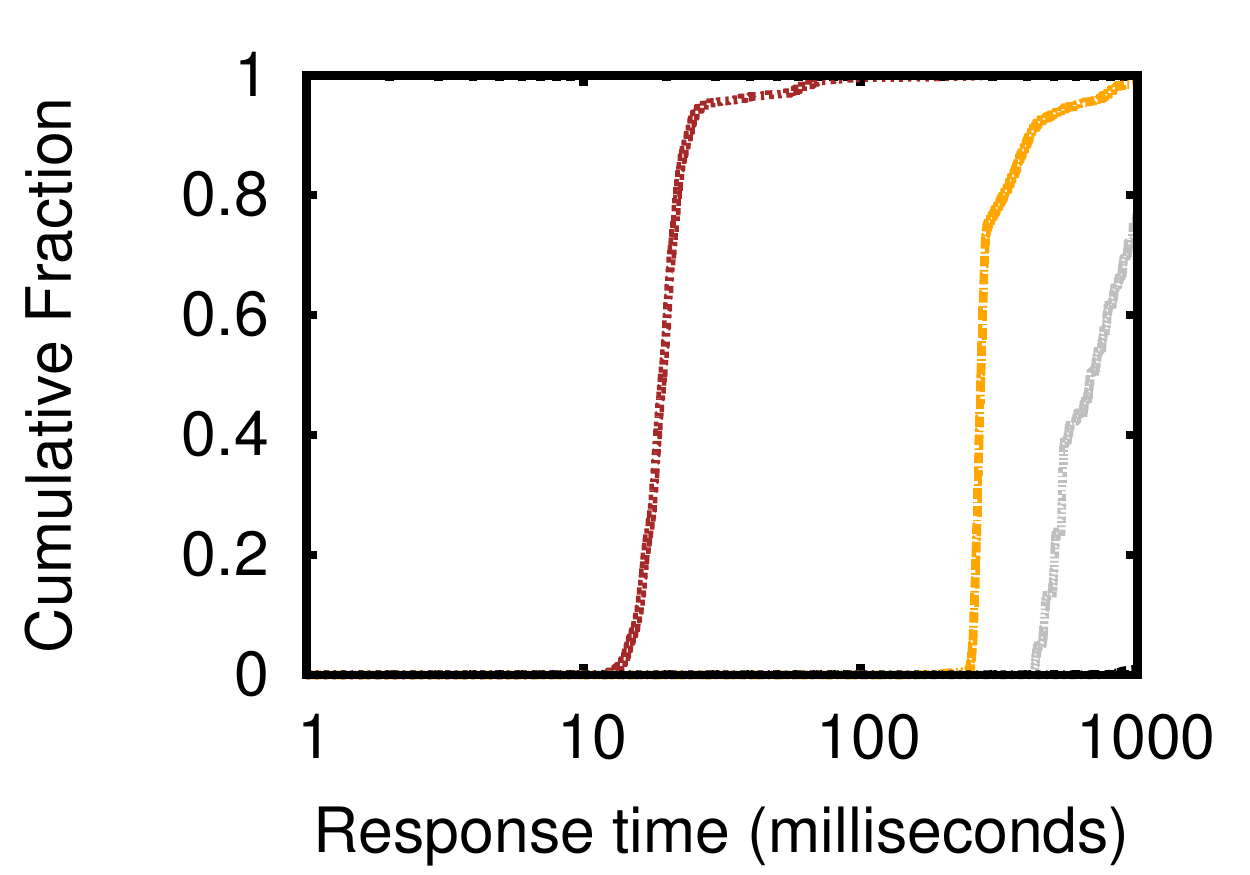}}
\subfigure[x86 with a single switch\label{fig:latency1x}]{\includegraphics[width=.79\columnwidth, clip=true, viewport=28 5 355 250]{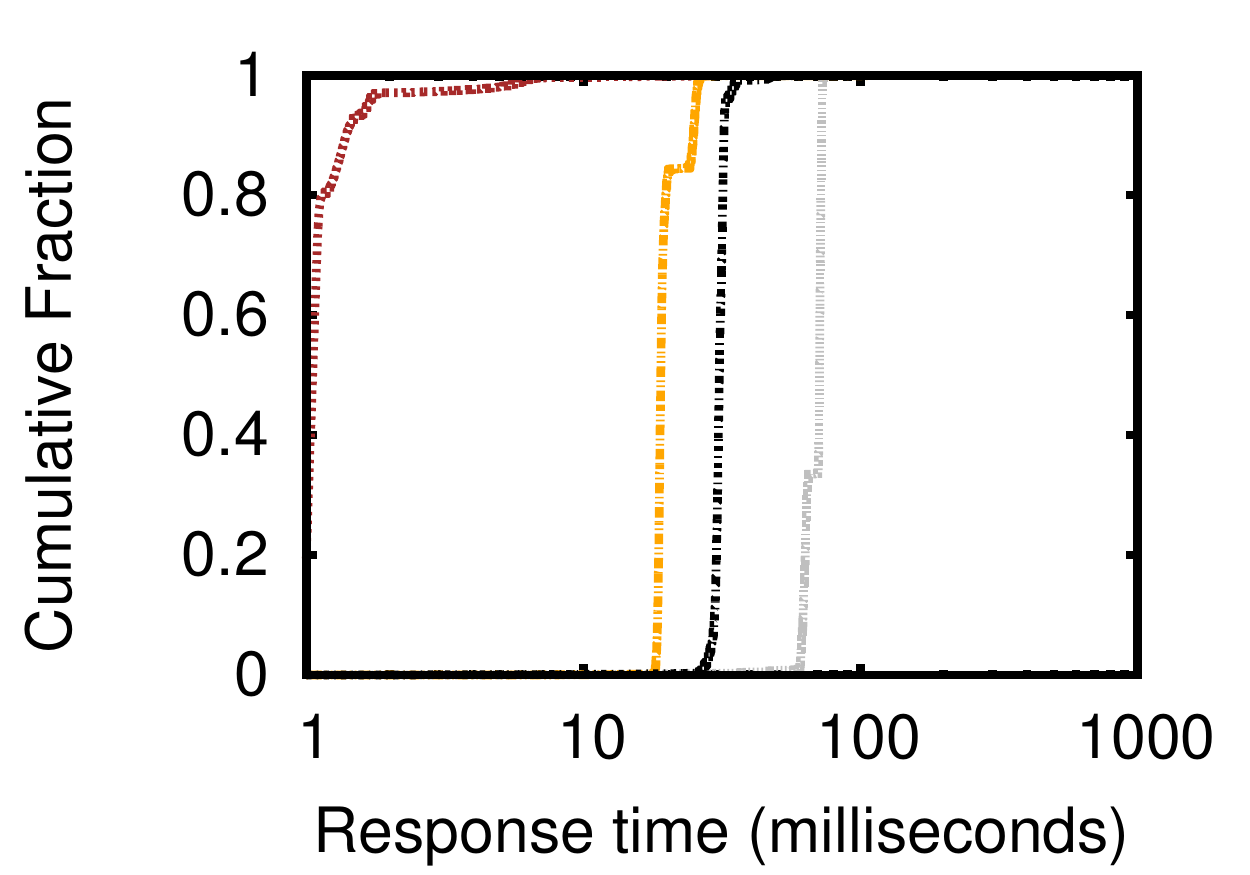}}\\
\subfigure[TileGx with 64 switches\label{fig:latency64t}]{\includegraphics[width=.79\columnwidth, clip=true, viewport=28 5 355 250]{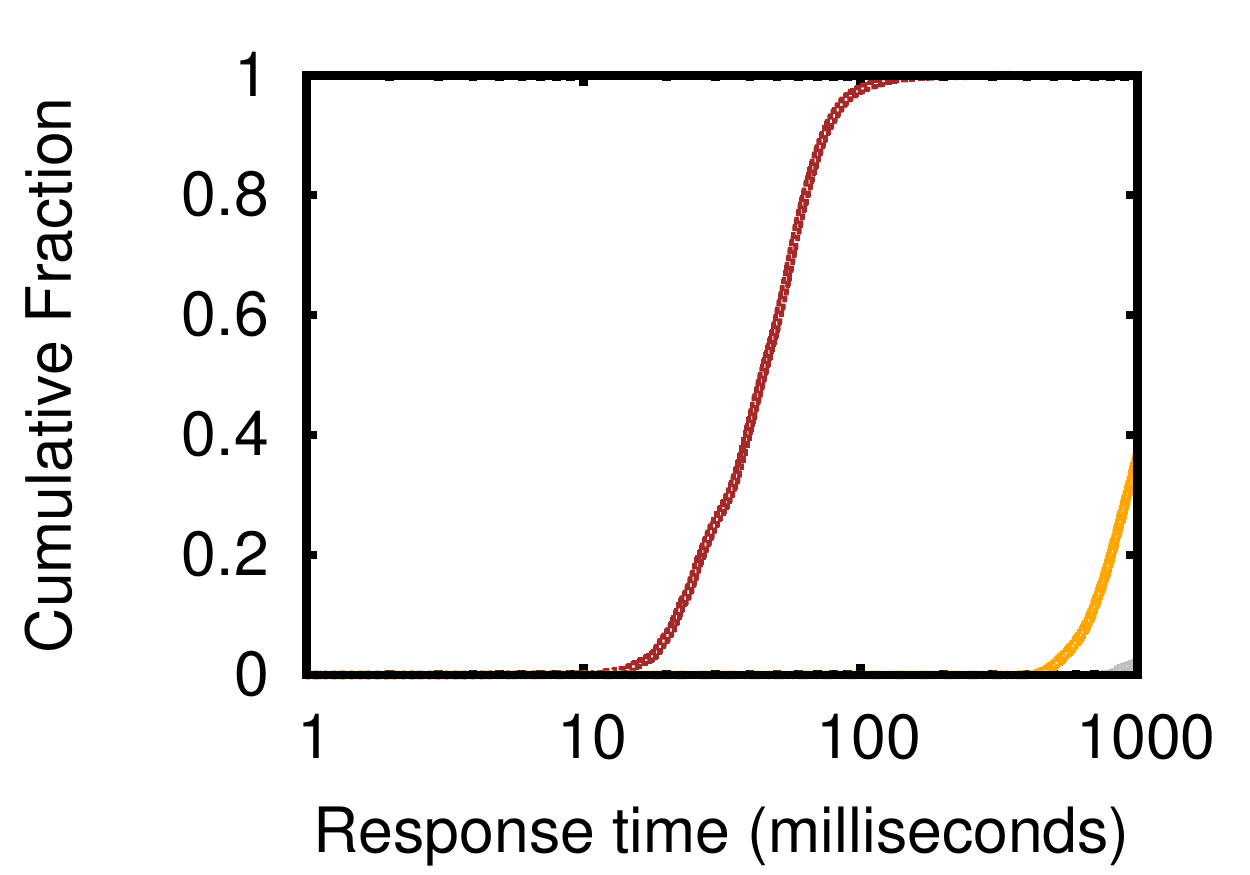}}
\subfigure[x86 with 64 switches\label{fig:latency64x}]{\includegraphics[width=.79\columnwidth, clip=true, viewport=28 5 355 250]{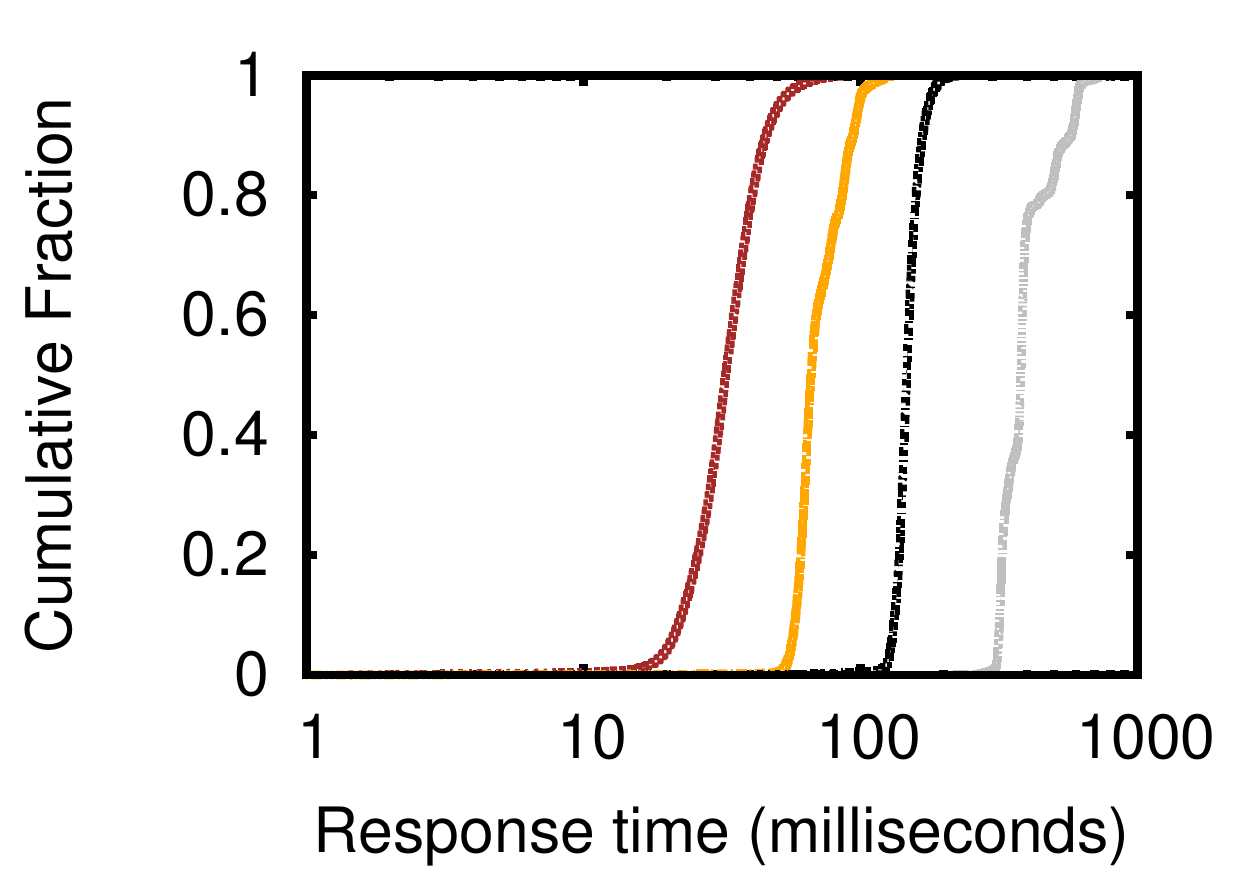}}\\
\subfigure{\includegraphics[width=.75\columnwidth]{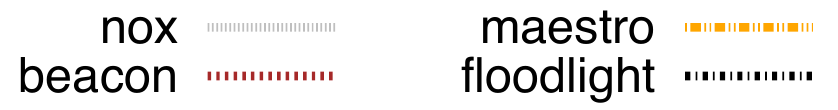}}
\caption{CDF of the latencies of SDN controllers on many-core (left) and multi-core (right) with one (top) and 64 switches (bottom)}
\label{fig:latency}
\end{figure*}

Figure~\ref{fig:latency} depicts the cumulative distribution function (CDF) of the response latency between switches and the controller installed on both platforms. In particular, in Figures~\ref{fig:latency1t} and \ref{fig:latency1x} we present the best case scenario when only 1 switch is connected to any given controller while Figures~\ref{fig:latency64t} and~\ref{fig:latency64x} present the more realistic scenario where 64 switches compete for the controller. 

On the multi-core platform, we observe in Figure~\ref{fig:latency1x} that while all controllers respond within $100$ms only Beacon responds within $10$ms which makes it acceptable for very short lived flows. In Figure~\ref{fig:latency64x}, we show that in the case of 64 switches only Maestro and Beacon respond within $100$ms and none respond within $10$ms. 

On the many-core platform, 
the latency is significantly higher. 
In fact, in the single switch setup only Beacon responds to a majority of the flows within $20$ms while the other three controllers struggle to respond within $1$ second. In the 64-switches setup, once again only does Beacon respond within $100$ms for the majority of the flows but more interestingly, the three others do not even respond within $1$ second.

\subsection{Energy Consumption}

\begin{figure}
\hspace{-1em}\includegraphics[angle=270,width=.5\textwidth]{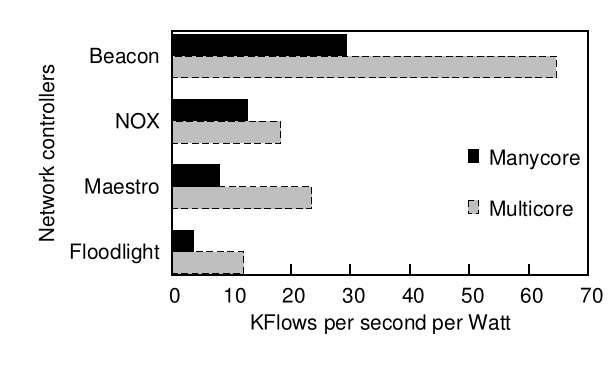}
\caption{Throughput per Watt of controllers on multi-core and many-core platforms}
\label{table:prof}

\end{figure}

Even though the raw performance on many-core is lower than on multi-core, many-core have a notoriously lower carbon footprint~\cite{borkar:2011,berezecki-etal:2011}, hence we expected 
to obtain a higher performance per Watt ratio on the many-core platform than on the multi-core platform.
We measured the energy consumption of each machine when stress testing each controller and observed  a peak energy consumption of 150W and 320W for the many-core and multi-core platforms, respectively. Based on these measurements, we present in  \figurename~\ref{table:prof} the throughput per Watt of the various controllers depending on the platform. 
It is clear that none of the four controllers exploits our many-core platform as each obtain a higher 
performance over energy ratio on the 32-way multi-core platform than on our 36-way many-core platform.

\section{Reasons of Limitations}
\label{sec:discussion}
In this section, we list the reasons of the throughput, latency and energy efficiency results obtained in the previous section by profiling each controller.

\subsection{Packet Serialization and Caching} 
The first limitation common to the four state-of-the-art controllers is their object-orientation.
Because their internal design 
treats OpenFlow packets as objects, 
messages must be parsed,  converted into objects and serialized for writing.
These steps 
increase the cache footprint. 
More dramatically, they induce a per-packet 
overhead, due to memory copying and allocations.

In particular, we found through profiling on the TileGx that NOX, Beacon and Maestro spend a significant amount of time in parsing and serializing. Interestingly, we found that Floodlight overhead for message parsing and serialization is not very important due to a lack of efficiency in the learning switch application as we will discuss later. 
On a many-core platform NOX spends about 46\% of its time invoking memory copies as depicted in Table~\ref{table:legacyprofiling}, while Maestro spends about 12\% serializing and sending packets and Beacon spends 13\% of its execution parsing packets into objects.
On highly efficient multi-core platforms where controller performance are typically limited by the memory bandwidth, the extra number of 
copies translates into further performance degradation.

\begin{table*}[t]
\centering
\caption{Profiling of the controllers running on many-core\label{table:legacyprofiling}}
\setlength{\tabcolsep}{26pt}
\begin{tabular}{lcccc}

\toprule
Description & NOX (\%) & Beacon (\%) & Maestro (\%) & Floodlight(\%) \\
\midrule
Memcpy & 45.99 & - & - & - \\
Learning switch application & 5.87 &  24.69 & 4.38 & 1.68\\
Hashmap accesses & 3.48 & 1.47 & 10.53& 6.06 \\
Packet output serialization & 2.12 & 21.16 & 11.80 & 3.46\\
Kernel &7.80 &14.84& 21.41& 31.31\\
\bottomrule
\end{tabular}
\end{table*}

All four controllers do not have any spatial locality awareness, data can be cached in a remote core to the thread that is using that data.
On the TileGx architecture, we observed a 12\% and 17\% remote read miss in both parsing and serializing packets for respectively Beacon and Floodlight, incurring additional latency due to network on chip traversals or remote cache accesses.
In fact, x86 cores share access to a large centralized L3 cache whereas TileGx cores have a distributed one. This results in additional overheads if requests need to fetch the cache of a remote core.

\subsection{Threading Model for Input/Output} 
The second limitation is the IO bottleneck common to Floodlight, Maestro and NOX even though they handle IO with different threading models.
Maestro has a single thread responsible for handling IO and splitting packets, this design does not scale to a large number of cores.
This overhead is more evident on many-core platforms whose clock frequencies are lower than on multi-core ones: the IO thread cannot provide packets fast enough for worker threads.

NOX uses the Boost.Asio library for handling IO, a single IO-service is run from a pool of worker threads.
This limits a single core performing the ``epoll'' system call
at once and adds overhead through the use of work and result queues, which are shared between worker threads.
The synchronization overhead from this threading model also limits the performance across multiple processor sockets due to intersocket sharing.

\subsection{The Shared Data Structure Bottleneck} 
The third limitation is due to the key data structure of the learning switch application: the hash table. 
Existing
controllers experience either a locking overhead or a large cache overhead
, which increases the time needed to process each individual packet.
We observed that NOX, Maestro and Floodlight experienced over 20\% of cache misses in accessing the hash table data structure.
All three allow packets to be processed out of order and this requires locking. Unfortunately, locks add significant overhead 
as the number of threads increases.

Maestro uses a generic Java concurrent hash map, the use of generics adds storage overhead that results in higher cache misses as well as 
boxing/unboxing 
 overhead, which requires an additional pointer traversal to fetch the value. 
We noticed that the generic storage overhead in Maestro results in higher local cache misses on the TileGx compared with NOX and Beacon.

\begin{table}[b]
\centering
\caption{Hash table cache misses on many-core\label{table:misses}}
\setlength{\tabcolsep}{18pt}
\begin{tabular}{lcc}
\toprule
Controller & local read miss  & remote read miss  \\
\midrule
Beacon & 0 \% & 0.15 \% \\
Maestro & 18.52 \% & 12.54 \% \\
NOX & 2.78 \% & 19.35 \% \\  
Floodlight&  6.17 \% & 13.89 \% \\
\bottomrule
\end{tabular}
\end{table}

NOX uses the \opfont{unordered\_map} from the Boost library that
resolves collisions using chaining by storing values with the same keys to a bucket implemented as a linked list.
The lists induce memory fragmentation and
greater storage overhead. Parsing this list involves indirections which leads, in turn, to 
poorer cache performance compared to the use of open addressing.
Profiling indicates that 20\% of remote read misses, where data is fetched from another core's cache as depicted in Table~\ref{table:misses} occurred on the hash table. This is a result of poor cache efficiency due to greater memory fragmentation and storage overhead.

\subsection{Other problems.} 
Beacon allocates switch input and output buffers in the listening thread, which introduces additional latency if the memory node for the worker thread is different from the listening thread.
Allocations that occur on the listening thread will tend to proceed on the closest memory controller, this can impact performance on multisocket systems.
\vincent{Is this a benefit of general many-core architecture?}
Allocating memory closest to where it will be used by allocating from the worker thread can result in better memory placement and caching on the TileGx. \vincent{Do we have references to Figure~\ref{?}...}
Beacon and NOX do not explicitly pin threads to cores, pinning threads reduces CPU migrations and can improve cache efficiency.
NOX demonstrates significantly lower per-thread performance on both x86 and the TileGx. 
NOX also appears to scale poorly when having less than a 2:3 ratio of threads to switches as depicted in Figure~\ref{fig:final_tilera_threads}, this effect was not observed with Beacon.

\subsection{Discussion}
\label{sec:legacydiscussion}

NOX demonstrates significantly lower per-thread performance on both x86 and the TileGx.
NOX also scales poorly when having less than a 2:3 ratio of threads to switches, this effect was not observed with Beacon.
NOX experiences overhead in making unnecessary copies of memory and synchronizing shared data structures such as the network view hash table for each switch.
The use of the Boost ASIO library also adds significant overhead through its use of task and completion queues as well as the limitation that only a single thread can call `epoll' at once.

Maestro demonstrates relatively good performance characteristics on x86, unlike NOX and Beacon it is able to scale when the number of threads exceeds the number of switches.
Maestro has poor per-thread performance on the TileGx and has limited scalability beyond 16 threads.
We consider two major reasons for this scalability as a single thread performing IO and the overhead of task queue synchronization.
The single thread performing IO is sufficient for x86 where cores are significantly more powerful, however this becomes a bottleneck on the TileGx.
Task queue and switch data structure synchronization also contributes to this poor performance. 

Beacon performs better than NOX and Maestro, but overall has lower performance than on x86.
The simpler IO design of Beacon results in lower overhead per-thread as well as better scalability on both processors.
The Beacon design minimizes contention by keeping switch data local to the processing thread. 

We have evaluated NOX, Beacon and Maestro on a many-core platform and compared their performance and scalability to an x86 processor.
We observe NOX reaches a similar maximum performance on both platforms but scales relatively poorly when the number of switches is greater than the number of cores.
Maestro scales relatively well on x86 but extremely poorly on the TileGx.
Beacon achieves the best performance on both platforms, having slightly better performance on x86.

Both NOX and Maestro have worse performance due to the overhead of synchronizing switch data structures.
Maestro's use of a single IO thread does not scale, particularly given the weaker performance of each core on the TileGx.
The simpler multi threaded design of Beacon through use of static switch partitioning to minimise synchronization results in better scalability on the TileGx.
Beacon fails to take advantage of the underlying hardware, the powerful memory management of cache control features of the TileGx could improve spatial locality of caching.

\section{Related Work}
\label{sec:related}

As far as we know, SDN controllers have never been ported before to manycore architecures, 
so we first present the 
existing performance results of concurrent SDN controllers on traditional x86 architectures and then discuss the benefit of manycore architectures observed with other network applications. 

\subsection{Performance of SDN Controller on x86 architectures}
Many controllers were proposed over the last years to exploit the concurrency of modern multi-core machines. Four of them in particular received an intensive attention, namely NOX~\cite{gude:2008, tootoonchian:2012}, Beacon~\cite{erickson:2013beacon},  Maestro~\cite{ngmaestro}, and Floodlight~\cite{erickson:2012floodlight}.

NOX-MT~\cite{tootoonchian:2012} is a multithreaded implementation of NOX. 
This advanced version uses asynchronous I/O provided by the C++ boost Asynchronous IO library (Boost ASIO) to simplify multithreaded operation.  
When profiling NOX, the original version of NOX spent 80\% of processor time on sending packets individually. 
Overall, NOX-MT outperformed NOX by a factor of 33 on a server with 2 quad core 2GHz processors~\cite{tootoonchian:2012}.

Maestro was the first controller to show performance scalability with the level of concurrency. It is written in Java and aims at dealing with unbalanced switch workloads~\cite{ngmaestro}. 
This approach binds switches to threads and results in better performance by reducing cache synchronization overhead. 
Maestro was evaluated on a 2 quad core AMD Opteron 2393 processors with one core 
dedicated to receiving requests and populating the task queues, as well as running the JVM garbage collection, leaving 7 cores for worker threads, 
hence achieving a maximum throughput of 0.6 Mreq/sec.

Beacon is another Java-based SDN controller.
Its author evaluated two different packet handling mechanisms called ``shared-queue'', similar to Maestro's task queue and ``run-to-completion''~\cite{erickson:2013beacon}. 
This evaluation was done through the deployment of controllers over an Amazon Elastic Compute Cloud (EC2) instance containing 16 physical cores from 2x Intel Xeon E5-2670 processors. 
Overall, Beacon reached a peak performance of 12.8 Mreq/sec (using the loopback interface) as compared to NOX achieving 5.3Mreq/sec and Maestro~\cite{ngmaestro} reaching 3.5Mreq/sec. 

Finally, Floodlight~\cite{erickson:2012floodlight} is a multithreaded Java-based SDN controller. Historically, Floodlight was the first controller reaching beyond the research community to the industry with its implementation as the Big Switch's Big Network Controller. 
This controller is now at the heart of the ONOS initiative~\cite{berde:2014} in which it was shown to respond to certain network events with a latency of 116ms for the 99$^{th}$ percentile.

\subsection{Network Applications on Many-Core Architecture}
Facebook implemented a Memcached Key-Value store on a TilePro64 processor \cite{berezecki-etal:2011}. 
They found despite the lower clock speeds, this processor performed with 67\% higher throughput than low power x86 processors at comparable latency.
More importantly they showed that the TilePro64 processor handled three times as many transactions per watt as the x86 processors with the same memory footprint. While the original memcached threading model had been 
significantly redesigned in this case to exploit the manycore platform, manycores (and in particular the Tilera architecture) have recently proved to provide higher performance per watts than multicores on existing benchmarks~\cite{GG16}.

More recently, such manycore platforms were shown to be effective at treating network packets.
Suricata is an open source high performance network IDS. 
Tilera announced that they achieved 25\,Gbps of throughput using Suricata on a TileExtremeGX platform.
IDS require high network throughput.
This platform is well suited towards this application as it provides high network bandwidth of 160\,Gbps of Ethernet and 144 TileGx cores \cite{TileraSuricata:2012}. 
The port of Suricata to the TileGX makes use of several hardware specific features provided to improve performance~\cite{Jiang:2013}.
Groups of cores are split into pipelines, one core in this pipeline processes incoming packets and dispatches packets on the UDN to the next core.
Worker cores then process packets according to a subset of rules before passing the packet onto the next core. 
The TileGX provides a low level API for handling these packets using Direct Memory Access to the networking hardware, this approach bypasses the Linux kernel.
Suricata also makes use of a ``dataplane mode'' where a subset of cores are removed from the Operating System scheduling, this disables scheduler interrupts and context switches allowing a single thread to run uninterrupted on those cores.

\section{Concluding Remarks}
\label{sec:conclusion}

We presented the most extensive evaluation of SDN controllers to date.  
This performance evaluation shows that these controllers 
cannot  
exploit
concurrent architectures mainly 
due to the inadequacy between the way controllers are designed and the unexploited features of the tested platforms. 
In particular, we demonstrated the impossibility for four state-of-the-art controllers to leverage 
the energy efficiency and high network traffic capabilities of many-core platforms.

Reasons for such a poor performance are manyfold. First, we found that all the controllers suffered from a per-packet overhead as the internal object-oriented design treats incoming packets as objects. Second, we  found that three out of the four controllers were limited by their threading model for IO. 
Third, their shared hash table data structure acts as a bottleneck  
where threads contend, thus requiring off-chip communication on multi-core platforms or requiring too much of the memory bandwidth of many-core platforms.

To cope with these issues we thus argue for a radically new way of designing SDN controllers.
Controllers should rather treat arriving packets with pre-allocated buffers rather than new objects and should be aware of the hardware characteristics to limit cache misses on multi-core platforms or exploit the network-on-chip on many-core platforms.

\section*{Acknowledgments}
 
This research was supported under Australian Research Council's Discovery Projects funding scheme (project number 160104801) entitled ``Data Structures for Multi-Core''.
Vincent Gramoli is the recipient of the Australian Research Council Discovery International Award.

\begin{acronym}[MACHU]

\acro{api}[API]{Application Programming Interface}
\acro{mcnc}[MCNC]{M*-Core Network Controller}

\end{acronym}

\bibliographystyle{IEEEtrans}\bibliography{references}

\end{document}